Article type: Full Article

# RP-CARS reveals molecular spatial order anomalies in myelin of an animal model of Krabbe disease.

*Giuseppe de Vito[1,2], Valentina Cappello[2], Ilaria Tonazzini[3], Marco Cecchini[3], and Vincenzo Piazza[2,\*]*

*Corresponding Author: E-mail: vincenzo.piazza@iit.it

[1] NEST, Scuola Normale Superiore, Piazza San Silvestro 12, I-56127 Pisa, Italy
[2] Center for Nanotechnology Innovation @NEST, Istituto Italiano di Tecnologia, Piazza San Silvestro 12, I-56127 Pisa, Italy
[3] NEST, Scuola Normale Superiore and Istituto Nanoscienze-CNR, Piazza San Silvestro 12, Pisa 56127, Italy

**Abstract.** Krabbe disease (KD) is a rare demyelinating sphingolipidosis, often fatal in the first years of life. It is caused by the inactivation of the galactocerebrosidase (GALC) enzyme that causes an increase in the cellular levels of psychosine considered to be at the origin of the tissue-level effects. GALC is inactivated also in the Twitcher (TWI) mouse: a genetic model of KD that is providing important insights into the understating of the pathogenetic process and the development of possible treatments. In this article an innovative optical technique, RP-CARS, is proposed as a tool to study the degree of order of the $CH_2$ bonds inside the myelin sheaths of TWI-mice sciatic-nerve fibres. RP-CARS, a recently developed variation of CARS microscopy, is able to combine the intrinsic chemical selectivity of CARS microscopy with molecular-bond-spatial-orientation sensibility. This is the first time RP-CARS is applied to the study of a genetic model of a pathology, leading to the demonstration of a post-onset progressive spatial disorganization of the myelin $CH_2$ bonds. The presented result could be of great interest for a deeper understanding of the pathogenic mechanisms underlying the human KD and, moreover, it is an additional proof of the experimental validity of this microscopy technique.

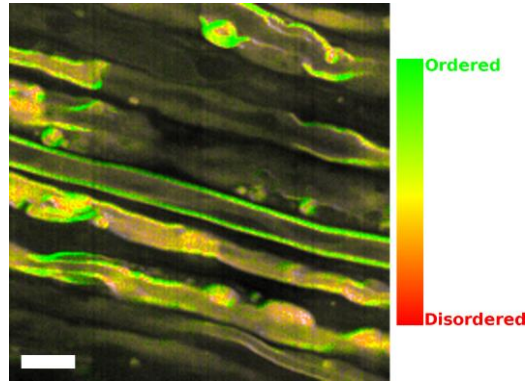

**Abstract Figure.** RP-CARS image (2850 cm$^{-1}$, CH$_2$ bonds) of a sciatic-nerve optical longitudinal section from a Twitcher P23 (symptomatic) mouse. Scale bar: 10 microns. The image was constructed by colour-mapping the degree of molecular order of the CH$_2$ bonds inside the myelin walls, as displayed in the colour bar on the right.

## 1. Introduction

Krabbe disease (KD), also called globoid cell leukodystrophy, is a rare (incidence: one in 100,000 births) genetic (autosomal recessive) sphingolipidosis caused by the inactivation of the galactocerebrosidase (GALC) enzyme [1]. GALC is a lysosomal enzyme implicated in the catabolism of galactolipids, including galactosyl-ceramide and psychosine (PSY, galactosyl-sphingosine). Its inactivation causes a cytotoxic increase in the cellular levels of PSY that is thought [2] to be at the origin of the tissue-level effects, even though the precise molecular mechanisms behind its cytotoxicity are still currently under discussion [3–13]. The increase of the PSY concentration in the central nervous system and in the peripheral nervous system leads to Schwann cell and oligodendrocyte death, causing progressive demyelination, associated with reactive astrocytosis, microgliosis [14] and neurodegeneration [15]. The onset is often early infantile [1] and, even though several promising treatments at the pre-clinical stage were proposed in the recent years [16–18], a cure for the human disease is still lacking and therefore KD is often fatal in the first years of life [19].

Twitcher (TWI) mouse is a genetic model of the human KD. It is an enzymatically authentic model of KD, since it is derived from a recessive spontaneous nonsense mutation of the GALC gene, and it closely resembles the histological, biochemical and morphological characteristics of human KD [20]. Microscopy techniques such as transmission electron microscopy [6,20], immunofluorescence [6,12,13,16–18] and confocal microscopy [6,13,16–18] have been used extensively to study the TWI mouse and they have provided important insights in the understating of the pathogenetic process and in the ideation of possible treatments. Recently a new optical imaging technique has emerged that promises to be an ideal tool for the study of myelin and its pathological states, namely Coherent Anti Stokes Raman Scattering (CARS) microscopy, thanks to the fact that its contrast generation mechanism is based on the presence of selected molecular bonds. Myelin displays an extremely high concentration of $CH_2$ bonds, comparable in the animal body only to adipose tissue or to some excretory glands.

The CARS process is an optical parametric process in which a couple of photons ("pump" and "Stokes") coherently excite a molecular vibration mode. A third photon ("probe") is used to probe the excited state. The system then relaxes to the ground state by emitting a fourth photon [21]. If the frequency difference of the incoming photons matches the frequency of the vibration mode, then the latter is resonantly excited and the generated signal intensity increases dramatically. CARS microscopy is a powerful and innovative technique that exploits the CARS process to image the sample at high spatial resolution without the use of chemical or biological probes, i.e. it is a completely label-free approach, based only on chemical contrast [22,23]. In this respect, it is similar to Raman micro-imaging but with orders-of-magnitude-larger signal levels, therefore affording faster acquisition times even in highly scattering (e.g. biological) samples [24].

Based on CARS imaging, a new polarization-resolved imaging technique named Rotating Polarization (RP) CARS [25,26] was recently developed. This technique exploits the CARS polarization-dependent rules [27] in order to probe the degree of anisotropy of the chemical-bond spatial orientations inside the excitation point-spread function (PSF) and their average orientation using a freely-rotating pump-and-probe-beam-polarization plane. In this technique the polarization-dependence of the CARS signal for each pixel is measured and analyzed in real time. It was already shown that the degree of anisotropy of the $CH_2$ bonds in myelin sheaths (quantified with a pixel-based numerical indicator) presents a significant correlation with their health status in a chemical model of demyelination [28]. In this article we shall report the use of this technique to visualize molecular spatial order anomalies in the genetic animal model of a leukodystrophy: the TWI mouse.

## 2. Materials and Methods

*2.1. Description of the RP-CARS microscope*

The RP-CARS microscope employed is fully described in [29]. For the sake of clarity, we summarize here its main components, as shown in **Figure 1**: an 810-nm pump-and-probe beam is generated by a mode-locked Ti:Sa laser ("fs Laser"; Chameleon Vision 2; Coherent Inc., Santa Clara, California, U.S.A.). After passing through a Faraday optical isolator ("Is"), 80% of it is used to pump an Optical Parametric Oscillator ("OPO"; Oria IR; Radiantis, Barcelona, Spain). The remaining 20%, after being further attenuated, plays the role of the CARS pump-and-probe beam. The Signal beam generated by the OPO at 1060 nm is used as Stokes beam (in order to excite the $CH_2$ vibration mode at 2850 cm$^{-1}$), while the Idler beam is dumped. The pump-and-probe and the Stokes beams are individually chirped by SF6 optical-glass blocks ("G" and "G'") and expanded by telescopic beam expanders ("BE" and "BE'"). The lengths of the SF6 blocks are precisely chosen in order to achieve spectral focusing [30] on the sample plane (20000 fs$^2$ of group-delay dispersion for both beams). The Stokes beam is made circularly polarized by means of a λ/4 retarder ("λ/4") while the polarization plane of the linearly-polarized pump-and-probe beam is continuously rotated by a brushless-motor-driven λ/2 rotating retarder ("R-λ/2"). An additional λ/4 retarder in the pump-and-probe path is used to compensate for the distortions induced by the optic elements (mostly by the dichroic mirrors) as well as a λ/2 retarder ("λ/2") in the Stokes path [27]. The two beams are synchronized by means of a delay line ("DL") present along the Stokes path and spatially recombined using a 900-nm long-pass dichroic mirror ("DM"). The beams are finally routed to a high-numerical-aperture (NA) objective ("Obj"; C-Achroplan W; 32X, NA=0.85, Carl Zeiss MicroImaging GmbH, Göttingen, Germany) of an inverted microscope (Axio Observer Z1; Carl Zeiss MicroImaging GmbH, Göttingen, Germany) through a pair of galvo-scanning mirrors ("XY"; GVS002; Thorlabs, Newton, New Jersey, U.S.A.), a scan lens, a compensation lens [29] and the tube lens of the microscope. The CARS signal generated from the sample ("S") is collected both in the trans direction, using a condenser lens ("Cond", NA=0.55), and in the epi direction, by means of a dichroic mirror ("DM'"), and then detected with red-sensitive photomultiplier tubes ("PMT"; R3896; Hamamatsu, Hamamatsu City, Japan) after being spectrally filtered in order to remove the pump photons and to select the Raman band of the $CH_2$ bonds (2850 $cm^{-1}$, "BP", band-pass filter centred at 650 nm and with a full width at half maximum of 50 nm).

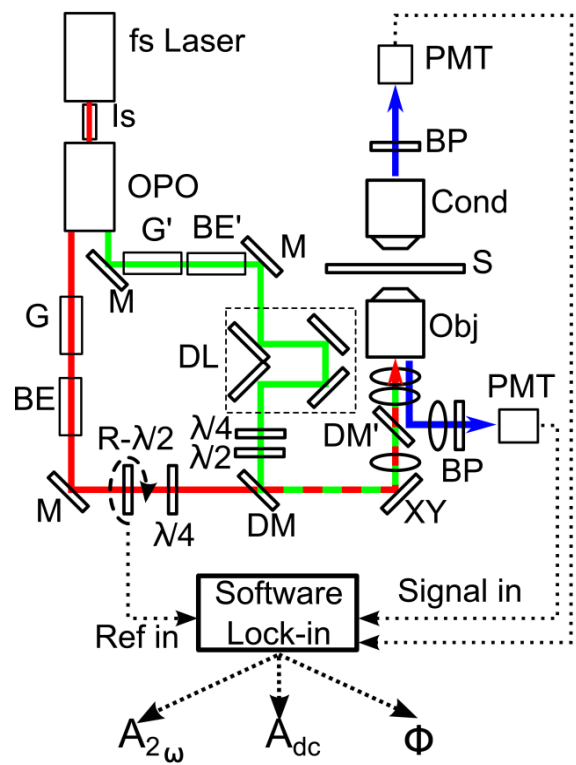

**Figure 1.** RP-CARS setup. Femtosecond laser (fs Laser), Faraday optical isolator (Is), optical parametric oscillator (OPO), SF6 optical glasses (G and G'), beam expander (BE and BE'), dichroic mirrors (DM and DM'), rotating half-

wave plate (R-λ/2), quarter-wave plates (λ/4), fixed half-wave plate (λ/2), delay line (DL), galvanometric mirrors (XY), sample (S), microscope objective (Obj), condenser (Cond), band-pass filters (BP), photomultipliers (PMT). The custom-made software (Software Lock-in) computes $A_{dc}$, $A_{2\omega}$ and $\Phi$ values from the detected signals. Thick red, green, and blue lines show the paths of the pump, Stokes and signal beams respectively. Figure adapted from: "G. de Vito, A. Canta, P. Marmiroli, and V. Piazza: A large-field polarization-resolved laser scanning microscope: applications to CARS imaging. Journal of Microscopy. 2015. 260. 194–199" [29]. Copyright Wiley-VCH Verlag GmbH & Co. KGaA. Reproduced with permission.

*2.2. Twitcher mice and sample preparation*

Twitcher heterozygous C57Bl/6J (B6.CE-Galc$^{twi}$/J) and wild-type (WT, C57Bl/6J) mice (Jackson Laboratory, Bar Harbor, Maine, U.S.A.) were generously donated by Dr. A. Biffi (San Raffaele Telethon Institute for Gene Therapy, Milan, Italy) and maintained under standard housing conditions. Twitcher heterozygous mice (Galc$^{twi/+}$) were used as breeder pairs to generate homozygous (Galc$^{twi/twi}$) TWI mice. Five wild type and seven homozygous TWI mice (aged from P15 to P38) were euthanized with cervical dislocation according to the ethical guidelines of the Italian (DLGS 26/2014) and European Community (2010/63/EU) laws. After death, the sciatic nerves were rapidly surgically explanted. Genomic DNA was extracted from the clipped tails of mice by Proteinase K lysis buffer as previously described [31] and the genetic status of each mouse was determined from the genome analysis of the twitcher mutation, as reported in [32].

The explanted nerves were put in WillCo dishes (GWSt-3522; WillCo Wells, Amsterdam, The Netherlands) filled with Krebs-Henseleit Buffer (K3753; Sigma-Aldrich, Saint Luis, Missouri, U.S.A.) and kept immobilized with an electrophysiology-type anchor (SHD-40; Warner Instruments, Hamden, Connecticut, U.S.A.). We acquired 3.5 z-stacks on average for each nerve.

*2.3. Signal acquisition and alpha-value algorithm*

The RP-CARS acquisition process is described in detail in [26]. Briefly, one galvanometric mirror scans repeatedly for a given number of repetitions (50 to 300 typically) an image line while the λ/2 retarder on the pump beam path rotates at ~10 Hz for the data presented here. When a line is completed, the system starts acquiring the following one while a custom-made software (developed in LabView; National Instruments Corporation, Austin, Texas, U.S.A.) reconstructs the temporal variation of the CARS signal for each pixel of the previous line and performs a pixel-based lock-in-like phase (Φ) and amplitude ($A_{2\omega}$) retrieval of the signal component at twice the rotation frequency of the polarization plane (20 Hz), together with the dc component ($A_{dc}$).

A second custom-made software (written in Python; Python Software Foundation, Beaverton, Oregon, U.S.A.) is used to compute the pixel-based alpha (α) value in post-processing as described in [28]:

$$\alpha = \tan^{-1}\left(\frac{A_{2\omega}}{A_{dc}}\right) \qquad (1)$$

This software also computes the average alpha (A) value for each z-stack acquisition in the following way: first the optical slice along the z-axis with the highest average $A_{dc}$ value is identified. Then the isodata algorithm [33] is applied to this slice in order to find the $A_{dc}$ threshold value. After that, all the pixels in the z-stack with $A_{dc}$ value higher than the threshold are selected and the average of their α values is taken as the A value of the z-stack. We used the highest threshold value that satisfies the Ridler-Calvard equation but, as a control, we also checked that different choices, for example using the lowest valid threshold value, do not lead to significant differences in the statistical analysis. We just observed a rather homogenous shift in the A values when different thresholding methods are chosen.

*2.4. Statistical analysis*

All the statistical computations were performed with R software (R Foundation, Vienna, Austria). We considered the TWI mice P20 or younger as (motor) presymptomatic phenotype, in accordance with the literature [20]. We averaged the A values derived from the z-stacks acquired from the same nerve and assigned them a weight equal to the inverse of their standard error. We used the averaged A values (and their weight) as the dependant variable in a General Linear Model and we used the genotype ($Galc^{twi/twi}$ or $Galc^{+/+}$) and the interaction between the genotype, the phenotype (symptomatic or presymptomatic) and the age (relative to the onset) as independent variable. Finally, we used Student t-test to compare the averaged A values between presymptomatic and symptomatic TWI mice.

*2.5. TEM imaging*

Late-pathological state TWI mice (P30 ± 2) under deep anaesthesia were sacrificed through perfusion of fixative solution (4% paraformaldehyde and 0.1% glutaraldehyde in phosphate buffer, pH 7.4). Sciatic nerves were dissected entirely and post-fixed for 4 hours at room temperature in the same fixative solution.

The samples were further treated for epoxy resin embedding. Briefly, the samples were deeper fixed in 2.5% glutaraldehyde in cacodylate buffer (0.1 M, pH 7.4). After rinsing, specimens were post-fixed with osmium tetroxide (1%) - potassium ferricyanide (1%) in cacodylate buffer, rinsed again, en bloc stained with 3% uranyl acetate in 20 % ethanol solution, dehydrated and embedded in epoxy resin, that was baked for 48 h at 60 °C. 80 nm-thick sections were obtained with an ultramicrotome (UC7; Leica Microsystems, Vienna, Austria) and collected on G300Cu grids (EMS, Hatfield, Pennsylvania, U.S.A.). Finally, sections were examined with a transmission electron microscope (Zeiss LIBRA 120 plus; Carl Zeiss MicroImaging GmbH, Göttingen, Germany) equipped with an in-column omega filter.

## 3. Results

*3.1. Qualitative analysis*

TWI-mouse phenotype is characterized by rapid myelin degeneration after the onset of the disease (P20). This is clearly seen in the **Figure 2** that depicts a large-field of view CARS image of a sciatic nerve of WT mouse compared to that one of a TWI mouse. In the TWI image the myelinated fibres appear sparser: this is because the pump and Stokes beams combination is chosen to target the $CH_2$ bonds and, in this condition, the demyelinated (or unmyelinated) fibres are invisible, lacking a sufficiently high $CH_2$ bonds concentration. Moreover, numerous alterations in the myelin-wall morphology are clearly visible.

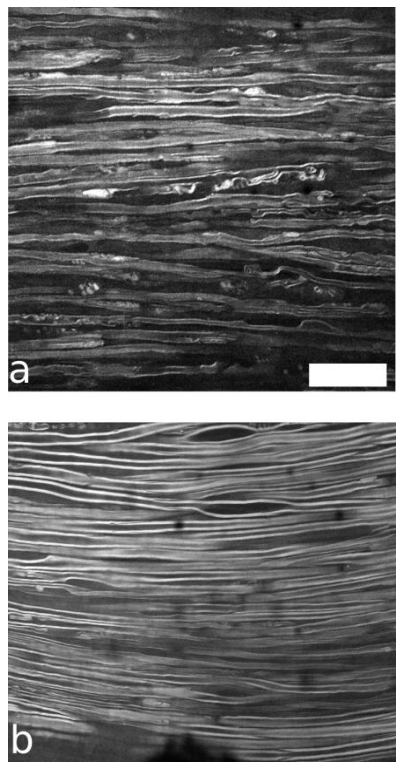

**Figure 2.** CARS images (2850 $cm^{-1}$, $CH_2$ bonds) of sciatic nerve optical longitudinal sections from a TWI P21 mouse **(a)** and of a WT P27 mouse **(b)**. Scale bar: 40 microns.

Higher resolution and magnification sciatic-nerve images are depicted in **Figure 3**. These images are representative of the WT, the presymptomatic TWI and the symptomatic TWI conditions. The hue in these images is constructed by colour-mapping the pixel-based α value, computed as described in the "Materials and methods" section. This value is a label-free optical indicator of the degree of molecular order of the $CH_2$ bonds inside the light excitation volume [25] and it is highly correlated to the myelin health status [28]. As displayed in this Figure, there is little difference between the WT and the TWI presymptomatic conditions (green colour), while it can be clearly seen that most of the myelinated fibres present ample portions of damaged myelin walls (yellow colour and tortuous appearance) in the TWI symptomatic condition. Unexpectedly, it is possible to occasionally observe long sections of myelin walls with almost normal morphological appearance in the symptomatic TWI condition images (indicated with white arrows in Figure 3C-D), even in the late period after onset (P25). It is worth noting that, despite their almost normal morphology, those fibres are associated with α values ranging from the values normally associated with WT and TWI presymptomatic fields (α ≈ 15, Figure 3C) to values similar to those of the other fibres in the respective field (Figure 3D).

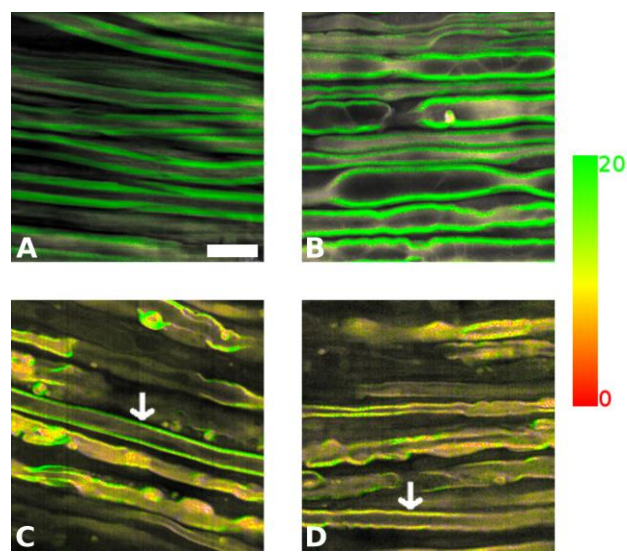

**Figure 3.** RP-CARS images (2850 cm$^{-1}$, $CH_2$ bonds) of sciatic nerve optical longitudinal sections from a WT P27 mouse **(A),** a TWI P20 (pre-symptomatic) mouse **(B)** and TWI P23 and P25 (symptomatic) mice **(C, D)**. Scale bar: 10 microns. The white arrows indicate sections of morphologically-normal myelin walls. The images were constructed in the HSV

(Hue, Saturation, Value) colour space by mapping the value of α onto the hue (from red, α = 0, to green, α = 20, as displayed in the colour bar on the right), the $A_{2\omega}$ signal onto the saturation and the $A_{dc}$ signal onto the value.

*3.2. Quantitative analysis*

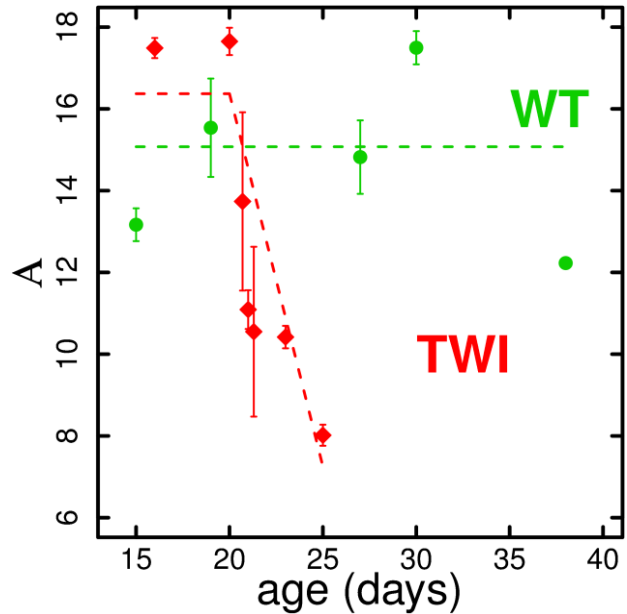

**Figure 4.** Graph of the average A values (on the y-axis) of WT (green, circles) and TWI (red, squares) mice of different ages (on the x-axis). The dashed lines indicate the predicted values by the statistical model. The p-value and the adjusted $R^2$ value of the statistical model were found to be 0.0012 and 0.7264 respectively. The error bars indicate the standard error of the A averages. X-axis jitter was applied to the P21 observations in order to improve the visualization.

For each z-stack we computed the A value by averaging the α values of the pixels associated to myelin, as described in the "Materials and methods" section; we constructed this value to be a global label-free optical indicator of the myelin health imaged in the stack. We then averaged the A values to obtain a single value for each animal and the results are shown in **Figure 4**. We found that before the onset age the A values of the TWI mice are constant, while after the onset they decrease in a linear manner (p = 0.0005, General Linear Model). In comparison, the A values of the WT mice remain constant over all the age range (model p-value = 0.0012, adjusted $R^2$ value = 0.7264).

In addition, we found the average A value of presymptomatic TWI mice being slightly higher (Δα ≈ 3) than that of the WT mice (p=0.03, Student t-test). This last result should be taken with caution, due to the small number of presymptomatic TWI mice, but it is important since it could suggest important implications about the structure of myelin in TWI mice before the onset of motor symptoms, as we explain in the "Discussion" session.

*3.3. TEM imaging*

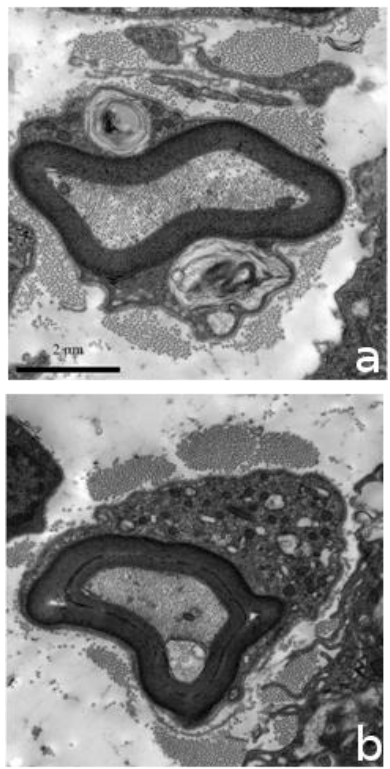

**Figure 5.** Representative images of Schwann cell wrappings in the sciatic nerve of P30 TWI mice. The deregulation of the staking could affect the outer portion of the nerve fibre **(a)** as well as the axonal side of myelin sheath **(b)**. The forthcoming death of myelinating cells appears also evident in the degeneration of cellular organelles, mainly mitochondria that appear vacuolated in the cytosol of Schwann cells.

The ultrastructural characterization of (P30 ± 2) TWI sciatic nerves, in agreement with CARS data, showed disorganization of the myelin sheaths architecture. As shown in Fig. 5, it is possible to observe within the sheath several points of discontinuity between subsequent windings (or wraps).

This effect is more evident in the periphery of the sheath, both in the inner and in the outer side. The final result of this alteration is highlighted by an increase of the myelin sheath thickness in comparison with control samples belonging to WT littermates.

## 4. Discussion

In this article we have shown that an optical technique (RP-CARS) is able to extract in real time information about structural arrangements in a fresh biological tissue (myelin) that are much below the physical limit for optical resolving power. Moreover, the combination of the RP-CARS observation and of the TEM observation gives useful insights on the degeneration of myelin during the after-onset period: our data suggest that the myelin degeneration involves, at least in the initial state, the contemporaneous appearance of numerous nanometric-sized (sub-optical resolution) structural alterations in the same myelin wall segment. We would like to underline that the label-free nature of the CARS imaging makes it possible to visualize the sample without the manipulations associated with fluorescent-labelling processes commonly needed for the fluorescence imaging, or with the tissue fixation and staining procedures required by TEM imaging, strongly minimizing the possibility of unwanted artefacts in the myelin microstructure.

A quite unexpected result was the sporadic observation (even in the late period after the onset) of normally appearing myelin walls associated with variable α values. Normally appearing myelin walls in TWI mice of similar ages were already observed [34] by means of TEM imaging and interpreted as remyelinated fibres. In the pursuit for a cure, it would be of great clinical interest the comprehension of the mechanisms behind the survival or the remyelination of those rare myelin walls while the nervous fibres around them are already completely demyelinised.

In addition, we can state that the myelin structure is probably altered at the nanoscopic level even before the clinical onset of the symptoms, when its microscopic morphological appearance seems otherwise normal. A possible explanation for this observation is the reduced expression of proteins, such as Myelin Basic Protein (MBP, the second most abundant protein in the myelin) in the juvenile period of TWI mice, as described in [35]. One of MBP roles is the adhesion of the cytosolic surfaces of compact myelin [36], therefore its reduction would be expected to lead to a general reduction of the intrinsic molecular order, however – to the best of our knowledge – there are no reported observations of decreased myelin interlamellar adhesion both in central [37] and in peripheral [34] nervous system of TWI mice of similar ages. On the other hand, the $CH_2$ bonds of the proteins are usually much less ordered with respect to those of the lipid-moiety of the myelin and this could be particularly true for intrinsically disordered proteins such as MPB [36], lacking an ordered three-dimensional structure. It could be hypothesized therefore that the reduction in the number of disordered protein-related $CH_2$ bonds leads to the observed global increase in the order of the myelin $CH_2$ bonds and this effect would overcompensates for the eventual presence of microstructural effects of the MBP deficit.

Finally, we would emphasize that RP-CARS has proven to be able to reliably produce an optical and label-free readout of the myelin health in an animal model of a leukodystrophy. Although KD is usually diagnosticated in a non-invasive manner, nerve biopsies are still required for the diagnosis of several other neuropathies, also involving myelin damage (e.g. chronic inflammatory demyelinating polyradiculoneuropathy [38], polyarteritis nodosa [39] or particular cases of Charcot–Marie–Tooth disease [40]). However, the surgical removal of a nerve sample causes a sensory deficit and may lead to chronic pain [41]. This limitation could be overcome by a technique able to assess the myelin health *in-vivo* without damaging the nerve (even though it would require an optical access). For this reason we speculate that the promising result presented here could open up in the future new possibilities for the development of less-invasive methods aimed at human neuropathies diagnosis.

## 5. Conclusions

We have exploited an innovative optical technique, RP-CARS, in order to demonstrate a post-onset progressive decrease in the spatial orientation order of the $CH_2$ bonds inside the myelin walls of TWI mice sciatic nerve fibres. This is the first time that this promising microscopy technique is applied to a genetic pathological model and our observations are an additional proof of the experimental validity of this microscopy technique. Finally, we believe that the obtained result might be of a great interest for a deeper understanding of the pathogenic mechanisms underlying the human Krabbe disease.

**References**

[1] National Library of Medicine. *Krabbe Disease*. August 2012. Web. June 2015. http://ghr.nlm.nih.gov/condition/krabbe-disease.

[2] K. Suzuki, Neurochem. Res. **23**, 251–259 (1998).

[3] V. Voccoli, I. Tonazzini, G. Signore, M. Caleo, and M. Cecchini, Cell Death Dis. **5**, e1529 (2014).

[4] A. B. White, M. I. Givogri, A. Lopez-Rosas, H. Cao, R. van Breemen, G. Thinakaran, and E. R. Bongarzone, J. Neurosci. **29**, 6068–6077 (2009).

[5] J. A. Hawkins-Salsbury, A. R. Parameswar, X. Jiang, P. H. Schlesinger, E. Bongarzone, D. S. Ory, A. V. Demchenko, and M. S. Sands, J. Lipid Res. **54**, 3303–3311 (2013).

[6] C. A. Teixeira, C. O. Miranda, V. F. Sousa, T. E. Santos, A. R. Malheiro, M. Solomon, G. H. Maegawa, P. Brites, and M. M. Sousa, Neurobiol. Dis. **66**, 92–103 (2014).

[7] G. Pannuzzo, V. Cardile, E. Costantino-Ceccarini, E. Alvares, D. Mazzone, and V. Perciavalle, Mol. Genet. Metab. **100**, 234–240 (2010).

[8] P. Formichi, E. Radi, C. Battisti, A. Pasqui, G. Pompella, P. E. Lazzerini, F. Laghi-Pasini, A. Leonini, A. Di Stefano, and A. Federico, J. Cell. Physiol. **212**, 737–743 (2007).

[9] S. Giri, M. Khan, R. Rattan, I. Singh, and A. K. Singh, J. Lipid Res. **47**, 1478–1492 (2006).

[10] E. Haq, S. Giri, I. Singh, and A. K. Singh, J. Neurochem. **86**, 1428–1440 (2003).

[11] M. Khan, E. Haq, S. Giri, I. Singh, and A. K. Singh, J. Neurosci. Res. **80**, 845–854 (2005).

[12] E. Haq, M. A. Contreras, S. Giri, I. Singh, and A. K. Singh, Biochem. Biophys. Res. Commun. **343**, 229–238 (2006).

[13] L. Cantuti-Castelvetri, E. Maravilla, M. Marshall, T. Tamayo, L. D'auria, J. Monge, J. Jeffries, T. Sural-Fehr, A. Lopez-Rosas, G. Li, K. Garcia, R. van Breemen, C. Vite, J. Garcia, and E. R. Bongarzone, J. Neurosci. **35**, 1606–1616 (2015).

[14] D. A. Wenger, M. A. Rafi, and P. Luzi, Hum. Mutat. **10**, 268–279 (1997).

[15] L. Cantuti Castelvetri, M. I. Givogri, H. Zhu, B. Smith, A. Lopez-Rosas, X. Qiu, R. van Breemen, and E. R. Bongarzone, Acta Neuropathol. **122**, 35–48 (2011).

[16] A. Ricca, N. Rufo, S. Ungari, F. Morena, S. Martino, W. Kulik, V. Alberizzi, A. Bolino, F. Bianchi, U. Del Carro, A. Biffi, and A. Gritti, Hum. Mol. Genet. **24**, 3372–3389 (2015).

[17] A. Lattanzi, C. Salvagno, C. Maderna, F. Benedicenti, F. Morena, W. Kulik, L. Naldini, E. Montini, S. Martino, and A. Gritti, Hum. Mol. Genet. **23**, 3250–3268 (2014).

[18] B. A. Scruggs, X. Zhang, A. C. Bowles, P. A. Gold, J. A. Semon, J. M. Fisher-Perkins, S. Zhang, R. W. Bonvillain, L. Myers, S. C. Li, A. V. Kalueff, and B. A. Bunnell, STEM CELLS **31**, 1523–1534 (2013).

[19] K. Suzuki, J. Child Neurol. **18**, 595–603 (2003).

[20] K. Suzuki and K. Suzuki, Brain Pathol. **5**, 249–258 (1995).

[21] P. D. Maker and R. W. Terhune, Phys. Rev. **137**, A801–A818 (1965).

[22] J.-X. Cheng and X. S. Xie, J. Phys. Chem. B **108**, 827–840 (2004).

[23] C. L. Evans and X. S. Xie, Annu. Rev. Anal. Chem. **1**, 883–909 (2008).

[24] C. L. Evans, E. O. Potma, M. Puoris'haag, D. Côté, C. P. Lin, and X. S. Xie, Proc. Natl. Acad. Sci. U. S. A. **102**, 16807–16812 (2005).

[25] G. de Vito, A. Bifone, and V. Piazza, Opt. Express **20**, 29369 (2012).

[26] G. de Vito and V. Piazza, Opt. Data Process. Storage **1**, 1–5 (2014).

[27] E. Bélanger, S. Bégin, S. Laffray, Y. De Koninck, R. Vallée, and D. Côté, Opt. Express **17**, 18419 (2009).

[28] G. de Vito, I. Tonazzini, M. Cecchini, and V. Piazza, Opt. Express **22**, 13733 (2014).

[29] G. de Vito, A. Canta, P. Marmiroli, and V. Piazza, J. Microsc. **260**, 194–199 (2015).

[30] T. Hellerer, A. M. K. Enejder, and A. Zumbusch, Appl. Phys. Lett. **85**, 25–27 (2004).

[31] P. W. Laird, A. Zijderveld, K. Linders, M. A. Rudnicki, R. Jaenisch, and A. Berns, Nucleic Acids Res. **19**, 4293 (1991).

[32] N. Sakai, K. Inui, N. Tatsumi, H. Fukushima, T. Nishigaki, M. Taniike, J. Nishimoto, H. Tsukamoto, I. Yanagihara, K. Ozono, and S. Okada, J. Neurochem. **66**, 1118–1124 (1996).

[33] T. W. Ridler and S. Calvard, IEEE Trans. Syst. Man Cybern. **8**, 630–632 (1978).

[34] J. M. Jacobs, F. Scaravilli, and F. T. De Aranda, J. Neurol. Sci. **55**, 285–304 (1982).

[35] B. Smith, F. Galbiati, L. Cantuti Castelvetri, M. I. Givogri, A. Lopez-Rosas, and E. R. Bongarzone, ASN NEURO **3** (2011).

[36] J. M. Boggs, Cell. Mol. Life Sci. CMLS **63**, 1945–1961 (2006).

[37] H. Takahashi, H. Igisu, and K. Suzuki, Acta Neuropathol. **59**, 159–166 (1983).

[38] J. P. Azulay, Rev. Neurol. **162**, 1292–1295 (2006).

[39] J. H. Stone, JAMA **288**, 1632–1639 (2002).

[40] D. Pareyson, Ital. J. Neurol. Sci. **20**, 207–216 (1999).

[41] C. L. Sommer, S. Brandner, P. J. Dyck, Y. Harati, C. LaCroix, M. Lammens, L. Magy, S. I. Mellgren, M. Morbin, C. Navarro, H. C. Powell, A. E. Schenone, E. Tan, A. Urtizberea, and J. Weis, J. Peripher. Nerv. Syst. **15**, 164–175 (2010).

**Graphical abstract.** A new microscopy technique (RP-CARS) is exploited to study a genetic model of the Krabbe disease, a fatal human demyelinating pathology. RP-CARS allows, in a completely label-free fashion, quantifying the molecular order within the axon myelin sheaths, highlighting the loss of myelin structural organization over sub-micrometric length scales during the disease progression. RP-CARS is a promising tool for the study of diseases affecting the structural organization of biological tissues.

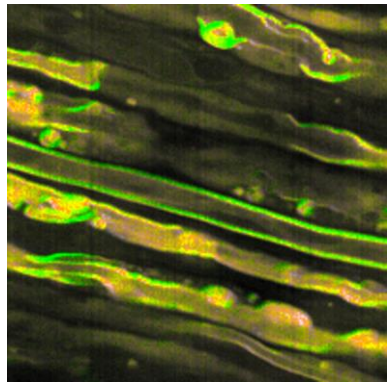